\documentclass[prl,floatfix,showpacs,aps,reprint,letterpaper,nobalancelastpage]{revtex4-1}
\usepackage{graphicx} 
\usepackage{color} 
\usepackage{transparent}
\usepackage{amsmath}
\usepackage{hyperref} 
\usepackage{amssymb}

\newcommand{\ket}[1]{\left\lvert #1 \right\rangle}

\begin{document}
\title{Microwave-activated conditional-phase gate for superconducting qubits}
\date{\today}

\author{Jerry M. Chow}
\affiliation{IBM T.J. Watson Research Center, Yorktown Heights, NY 10598, USA}
\author{Jay M. Gambetta}
\affiliation{IBM T.J. Watson Research Center, Yorktown Heights, NY 10598, USA}
\author{Andrew W. Cross}
\affiliation{IBM T.J. Watson Research Center, Yorktown Heights, NY 10598, USA}
\author{Seth T. Merkel}
\affiliation{IBM T.J. Watson Research Center, Yorktown Heights, NY 10598, USA}
\author{Chad Rigetti}
\affiliation{IBM T.J. Watson Research Center, Yorktown Heights, NY 10598, USA}
\author{M. Steffen}
\affiliation{IBM T.J. Watson Research Center, Yorktown Heights, NY 10598, USA}
\begin{abstract}
We introduce a new entangling gate between two fixed-frequency qubits statically coupled via a microwave resonator bus which combines the following desirable qualities: all-microwave control, appreciable qubit separation for reduction of crosstalk and leakage errors, and the ability to function as a two-qubit conditional-phase gate. A fixed, always-on interaction is explicitly designed between higher energy (non-computational) states of two transmon qubits, and then a conditional-phase gate is `activated' on the otherwise unperturbed qubit subspace via a microwave drive. We implement this microwave-activated conditional-phase gate with a fidelity from quantum process tomography of $\sim 87\%$. 
\end{abstract}
\maketitle

Superconducting qubits are a prime candidate for scaling towards larger quantum processors. Improvements to coherence times for Josephson junction-based qubits \cite{Paik2011,Corcoles2011,Rigetti2012,Chang2013} have made possible high-fidelity universal gates as well as the characterization of gate verification and validation methods~\cite{Chow2012, Magesan2012, Gambetta2012, Poletto2012, Corcoles2012arx, Merkel2012arx}. As systems evolve towards quantum error-correction architectures such as the two-dimensional surface code \cite{Bravyi1998,Raussendorf2007}, it is increasingly important to devise and characterize different entangling gate schemes to determine suitability for large-scale implementations. 

There have been multiple incarnations of universal entangling two-qubit gates for superconducting qubits, whether as $i\text{SWAP}$, controlled-NOT (CNOT), or conditional-Phase (c-Phase), each with their own set of advantages and disadvantages. One class of gates include all of those which rely on the dynamical flux-tunability of either the underlying qubits, or some separate  sub-circuit. This includes the direct-resonant $i\text{SWAP}$ (DRi) ~\cite{Bialczak2010,Dewes2012}, the higher-level resonance induced dynamical c-Phase (DP) \cite{dicarlo_2009,Strauch:2003ss,Yamamoto2010}, and any variant gates induced via a dynamic tunable coupling \cite{Niskanen:2007if,Bialczak2011}. Another class of gates contains all those in which the qubits have fixed-frequencies, and only a microwave-modulated passive coupling in place either directly or via a coupling circuit such as a resonator bus~\cite{majer_coupling_2007}. The gates in this class include the resonator sideband induced $i\text{SWAP}$ (RSi)~\cite{Leek:2009ht}, the cross-resonance (CR) gate which generates a CNOT \cite{Paraoanu2006,Rigetti2010,Chow2011,Corcoles2012arx}, the Bell-Rabi (BR) single-step entanglement gate \cite{Poletto2012}, the wait c-Phase (WP) gate from an always on $ZZ$ interaction~\cite{Poletto2013_inprep}, and the driven resonator induced c-Phase (RIP)~\cite{Gambetta2012aps,Paik_inpreparation2013}. 

The primary advantage of the dynamically tunable class of gates (DRi and DP) is the ability to operate the qubits in very different regimes: one in which the qubits are independent with negligible interaction, and one where the two-qubit interaction is maximized. In the first regime, single-qubit gates can be applied trivially without the need for specialized decoupling schemes as the qubits will not experience significant crosstalk errors. In the second regime, the qubits can be tuned to optimize the two-qubit interaction so as to enable the shortest possible gate times. This means simple single-qubit gates, the possibility of strong two-qubit interactions, and low crosstalk errors are enabled by DRi and DP. The main disadvantages of such gates are the reliance on flux-tunable qubits, which can have reduced coherence times due to flux-noise~\cite{yoshihara_decoherence_2006}, the risk of interacting with other energy levels in the system during tuning, and additional circuit and control complexity due to on-chip tunable flux controls or couplers which support dynamical tunability.

\begin{figure}[tbp!]
\centering
\includegraphics[width=0.47\textwidth]{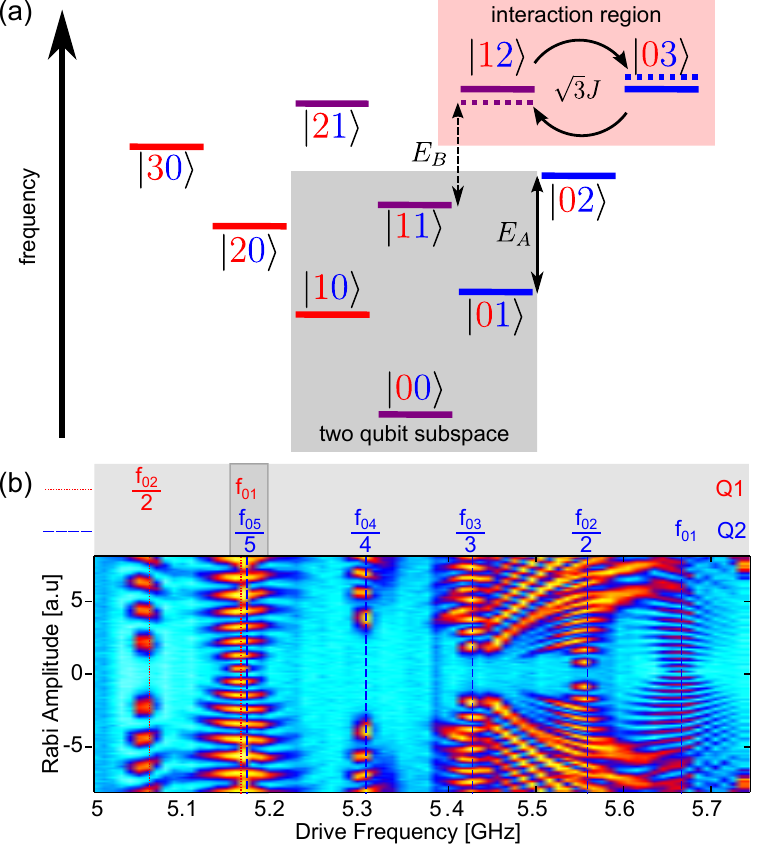}
\caption{\label{fig:1} (color online)~Level diagram and two-qubit Rabi amplitude spectroscopy. (a) Representative two-transmon energy ladder up to the three-excitation manifold. An interaction of strength $\sqrt{3}J$ between the $\ket{12}$ and $\ket{03}$ levels (upper right pink shaded box) gives rise to the MAP interaction, as it causes the energy difference between $\ket{11}$ and $\ket{12}$, denoted $E_B$, to be different from the energy difference between $\ket{01}$ and $\ket{02}$, denoted $E_A$. An off-resonant drive tone from $E_A$ will then result in a different phase on the $\ket{11}$ state relative to the other states in the computational basis states shaded in grey. (b) Density plot of Rabi amplitude spectroscopy. By varying the amplitude of a Rabi drive pulse and sweeping over frequency, it becomes possible to identify the full two-transmon energy landscape in cavity transmission. Starting from the right at higher frequencies and sweeping left, first the $f_{01}$=5.68 GHz transition of Q2 is encountered, which has a linear fringing pattern with increasing drive amplitude, subsequently followed by multi-photon transitions to the higher levels of Q2. Around 5.18 GHz, the $f_{01}$ of Q1 is observed, and it happens to overlap closely to the $f_{05}/5$ 5-photon transition of Q2. This observed resonance for these transmons which have identical anharmonicities, satisfies the condition necessary to observe the MAP interaction (exercise left for the reader).}
\end{figure}

For the case of the microwave two-qubit gates listed above, the qubits are fixed in frequency, and thereby can be parked at `sweet-spots' of coherence or made to be un-tunable from the start. Furthermore, the addressing hardware and shaped-microwave controls become analogous to those of single-qubit gates. There is additional circuit complexity for some of the schemes, specifically the RSi and CR gates require local microwave addressability for each qubit. The most significant disadvantages for the fixed-frequency gates are tradeoffs to either coherence or single-qubit control in order to have stronger two-qubit interactions. For RSi and RIP, the two-qubit gate interaction is optimized through having stronger resonator-qubit coupling strengths, $g$. Yet, inherent to both schemes is a step in which real photons exist in the cavity, and with large $g$, this can lead to significant dephasing due to photons during the operation \cite{gambetta_qubit-photon_2006}. As for CR, BR, and WP, the qubit-qubit detunings which would give the strongest two-qubit interaction, also happen to result in reduced single-qubit addressability, although this error can be mitigated via optimal-control schemes.

In this paper, we introduce an additional entangling gate to the lexicon of two-qubit gates: the microwave-activated c-Phase (MAP) gate. The MAP gate combines the higher-level resonant interactions introduced in the DP gate with the simple microwave controls of fixed-frequency transmons, while also permitting the transition between separate regimes for single-qubit control and two-qubit interaction, turned on or off via microwaves. The key to the MAP gate interaction lies in going beyond three levels of the transmon, and pre-defining a resonance condition in the three-excitation manifold. By designing transmons such that the $\ket{03}$ energy transition aligns with the $\ket{12}$ transition [$\ket{nm}$ refers to $n$ excitations in qubit 1 (Q1) and $m$ excitations in qubit 2 (Q2)], these levels are split by the inter-qubit interaction ($\sqrt{3}J$ in this manifold) and the degeneracy between the $\ket{02}\leftrightarrow \ket{01}$ and the $\ket{12}\leftrightarrow \ket{11}$ transition is removed. The net result is that an applied external drive near resonance with the $\ket{n2}\leftrightarrow \ket{n1}$ transition induces a $ZZ$ interaction, much like the ac-Stark effect. Here, we implement the MAP gate on a pair of transmon qubits coupled via a resonator bus designed with a detuning so as to align the $\ket{03}$ and $\ket{12}$ transitions. We characterize the interaction and tune-up a c-Phase gate which is verified via quantum process tomography (QPT) with a gate fidelity of 87\%. The technique is extendable to a general class of two-qubit MAP gates which can arise from other resonance conditions in the higher manifolds of transmon qubits. Furthermore, although the MAP gate significantly eases single and two-qubit controls, the burden of scaling this to larger systems lies in the fabrication of qubits with explicit resonance conditions in a well-defined window of energies.

The MAP scheme relies on the presence of higher levels in the two transmons, but unlike the DP gate, does not require any resonance condition between higher levels and computational states (i.e.~$\ket{00}$,\,$\ket{01}$,\,$\ket{10}$,\,$\ket{11}$). Rather, by careful control over the design of the transmons, through  capacitance and/or Josephson junction critical currents, it is possible to tailor the two different transmon energy spectra to experience a resonance condition involving only higher level non-computational states. 

In the case where the energy corresponding to $\ket{03}$ is aligned with the energy corresponding to $\ket{12}$ [see level diagram of Fig.~\ref{fig:1}(a)], the interaction between the transmons will result in a splitting of these levels by $2\xi$ where
\begin{eqnarray}
\xi = \frac{1}{2}\left(\sqrt{4J_{12,03}^2+\Delta_{12,03}^2}+\Delta_{12,03}\right).
\end{eqnarray} Here $J_{12,03}$ is the matrix element for the interaction between the 12 and 03 levels and $\Delta_{12,03}$ is the difference between the bare energies of the 12 and 03 levels. For a transmon, $J_{12,03}$ is approximately equal to $\sqrt{3}J$ and $\Delta_{12,03}$ is approximately equal to $\omega_1-\omega_2-2\delta_2$ where $\omega_1$ and $\omega_2$ are the 0-1 transition energies of the two transmons and $\delta_2$ is the anharmonicity of the second transmon.
As illustrated in Fig.~\ref{fig:1}(a) the transition energies between $\ket{11}$ and $\ket{12}$ [labeled $E_B$] and $\ket{01}$ and $\ket{02}$ [labeled $E_A$] differ by an amount $\xi$, i.e. $\xi$ can be viewed as a conditional anharmonicity. 

\begin{figure}[h!]
\centering
\includegraphics[width=0.47\textwidth]{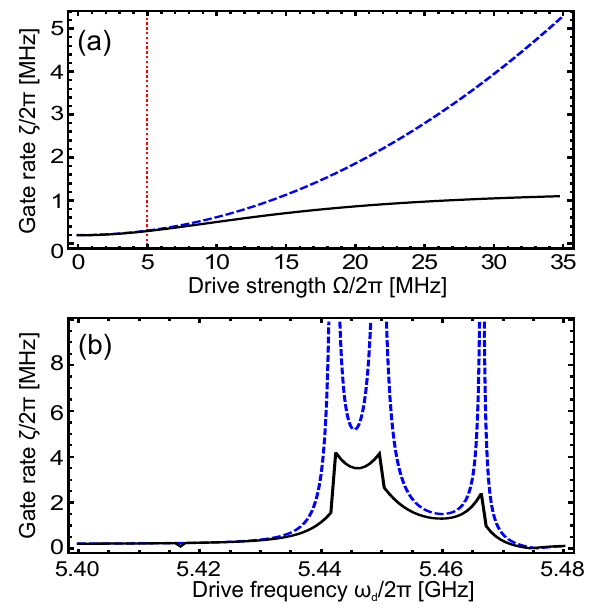}
\caption{\label{fig:x}(color online)~Perturbative theory. (a) At weak drive strengths (below $5$ MHz here), perturbation theory (dashed blue line) agrees with a six level numerical simulation (solid black line). Experiments are performed at higher drive strengths, where perturbation theory does not apply, and the gate rate of the numerical simulation is seen to saturate. Transition frequencies are taken from Rabi amplitude spectroscopy measurements [Fig.~\ref{fig:1}b] and drive frequencies correspond to those in Fig.~\ref{fig:3}. (b) At a drive strength of $5$ MHz, the perturbation theory (dashed blue line) agrees with numerics (solid black line) if the drive frequency is sufficiently far from a transition.}
\end{figure}

The basic principle of the MAP gate is to use this energy difference to induce a gate via the ac-Stark effect \cite{ct_atom_photon_interactions}. The ac-Stark shift of an energy level occurs when an external drive with amplitude $\Omega$ is nearly resonant with a transition involving that level. The level then shifts by an amount equal to the power of the external drive $\Omega^2$ divided by the difference $\Delta_d$ between the transition frequency and the drive frequency, provided that the ratio $\Omega^2/\Delta_d$ is small. At second order in perturbation theory, we find that the conditional-phase gate has a rate
\begin{equation}
\zeta=(E_{11}-E_{01}-E_{10})/\hbar\approx\zeta_0+\frac{\Omega^2}{2\Delta_d}\zeta_2
\end{equation}
where $\Delta_d=\Delta_{12,11}-\omega_d$. Here
\begin{equation}
\zeta_0 = J_{11,20}J_{11,02}\left(\frac{1}{\Delta_{11,20}}+\frac{1}{\Delta_{11,02}}\right)
\end{equation}
is the always-on component of the rate, and
\begin{equation}
\zeta_2 = \frac{J_{12,03}^2}{J_{12,03}^2+\Delta_d(\omega_d-\Delta_{03,11})}
\end{equation}
is the microwave-activated component.
While this is a reasonable approximation in the small drive limit [see Fig.~\ref{fig:x}], we find that as the drive strength is increased numerical simulations predict a saturation of this rate. Understanding this saturation will be a topic of future research.

\begin{figure}[h!]
\centering
\includegraphics[width=0.47\textwidth]{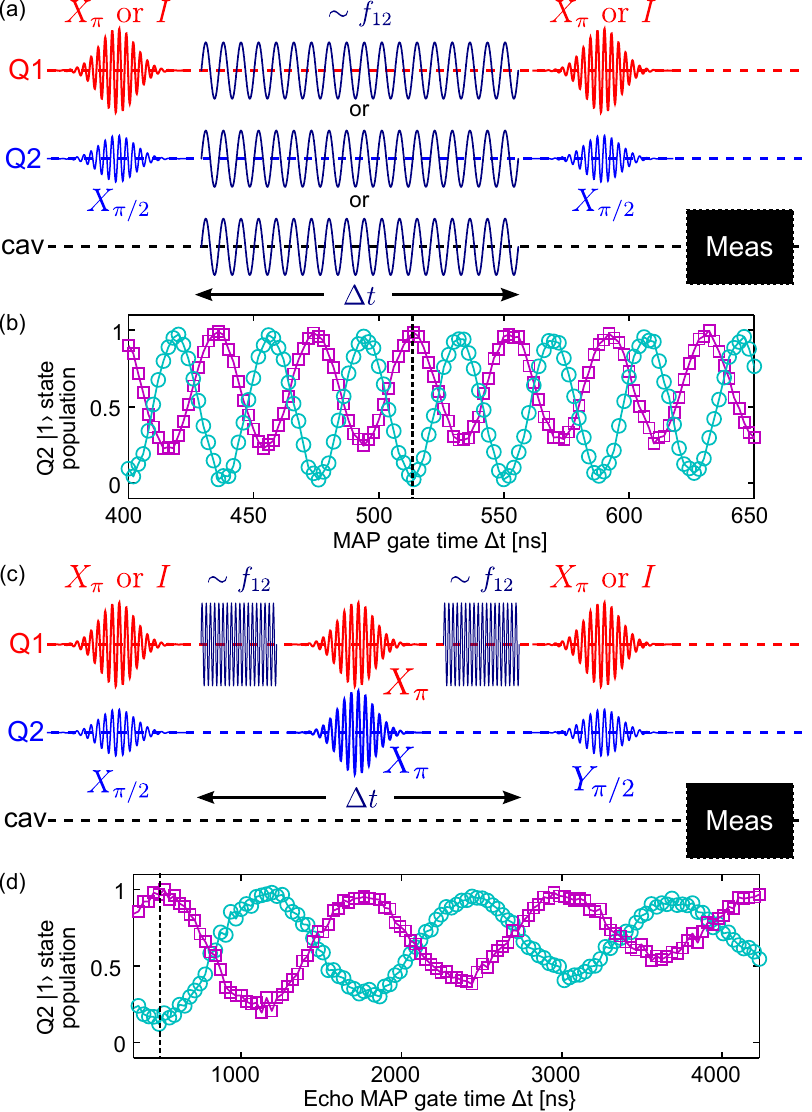}
\caption{\label{fig:2}(color online)~MAP interaction tune-up. (a) Pulse protocol for the simple MAP gate tune-up, involving a standard  Ramsey-type experiment, where the phase on Q2 is observed via time-separated $X_{\pi/2}$ pulses (gate length 40 ns), in the two cases of Q1 in ground or excited state. The MAP interaction is tuned via a microwave pulse tuned near $f_{12}$ of Q2 of varying amplitude and duration $\Delta t$ applied on any drive line. (b) The population of the $\ket{1}$ state for Q2, show different Ramsey fringes for starting in $\ket{00}$ (cyan circles) versus $\ket{10}$ (purple squares), and a $[ZZ]_{\pi}$ gate occurs at 514 ns (dashed line). (c) A modified cross-resonance, refocused MAP gate protocol which includes a composite pulse of total time $\Delta t$ consisting of three parts: two drive pulses on Q1 at an activation frequency near $f_{12}$ of Q2, separated by refocusing pulses $X_{\pi}$ on both Q1 and Q2. The Ramsey experiment is modified with a $Y_{\pi/2}$ at the end to observe out-of-phase fringes (d) controlled on the state $\ket{00}$ (cyan circles) or $\ket{10}$ (purple squares). A $[ZZ]_{\pi}$ gate is indicated at the first out-of-phase fringe (dashed line) at a total gate time of 510 ns. Note that the upwards trend of the Q2 $\ket{1}$ state population is a result of relaxation, as in the actual experiment protocol an additional $X_{\pi}$ is applied at the end to both qubits to undo the effect of the refocusing $X_{\pi}$ pulses.}
\end{figure}

For our experiment, we design two transmons ($\omega_1/2\pi$=5.166 GHz and $\omega_2/2\pi$=5.668 GHz) with a detuning close to twice the anharmonicity, $\delta_2/2\pi = -220$ MHz. Fig.~\ref{fig:1}(b) shows a Rabi amplitude spectroscopy landscape, where a strong Rabi-drive pulse of varying amplitude is applied to the coupling cavity ($\omega_r/2\pi$=8.646 GHz) of two transmons. Fringes are observed for each transition from the ground state to the $n$th level of transmon Q1 or Q2 [labeled $f_{0n}$ for each transmon in Fig.~\ref{fig:1}(b)]. The two transmons shown satisfy the MAP condition from the observation that the $f_{01}$ transition of Q1 aligns with the $f_{05}/5$ five-photon transition of Q2, which is generally $2\delta_2/2\pi$ detuned~\cite{koch_charge-insensitive_2007} from the $f_{01}$ transition of Q2.

The MAP interaction is tuned via a Ramsey experiment on Q2 as illustrated in Fig.~\ref{fig:2}(a). Conditioned on the state of Q1, different Ramsey fringes [Fig.~\ref{fig:2}(b)] are observed when a drive $\Omega$ at $\omega_{\text{d}}/2\pi\sim f_{12}$ of Q2 is applied to the system. A two-qubit c-Phase gate generator $[ZZ]_{\pi}=\text{exp}(-i\frac{\pi}{4}(Z\otimes Z))$ is realized when the fringes are $\pi$ out of phase, indicated by the dashed line at a gate time $\Delta t$ = 514 ns. Although the data in Fig.~\ref{fig:2}(b) is for $\Omega$ applied directly to Q2, the interaction can also be driven through the bus resonator, or in a CR-like scheme via the Q1 excitation-port, as indicated in Fig.~\ref{fig:2}(a). 

\begin{figure}[b!]
\centering
\includegraphics[width=0.47\textwidth]{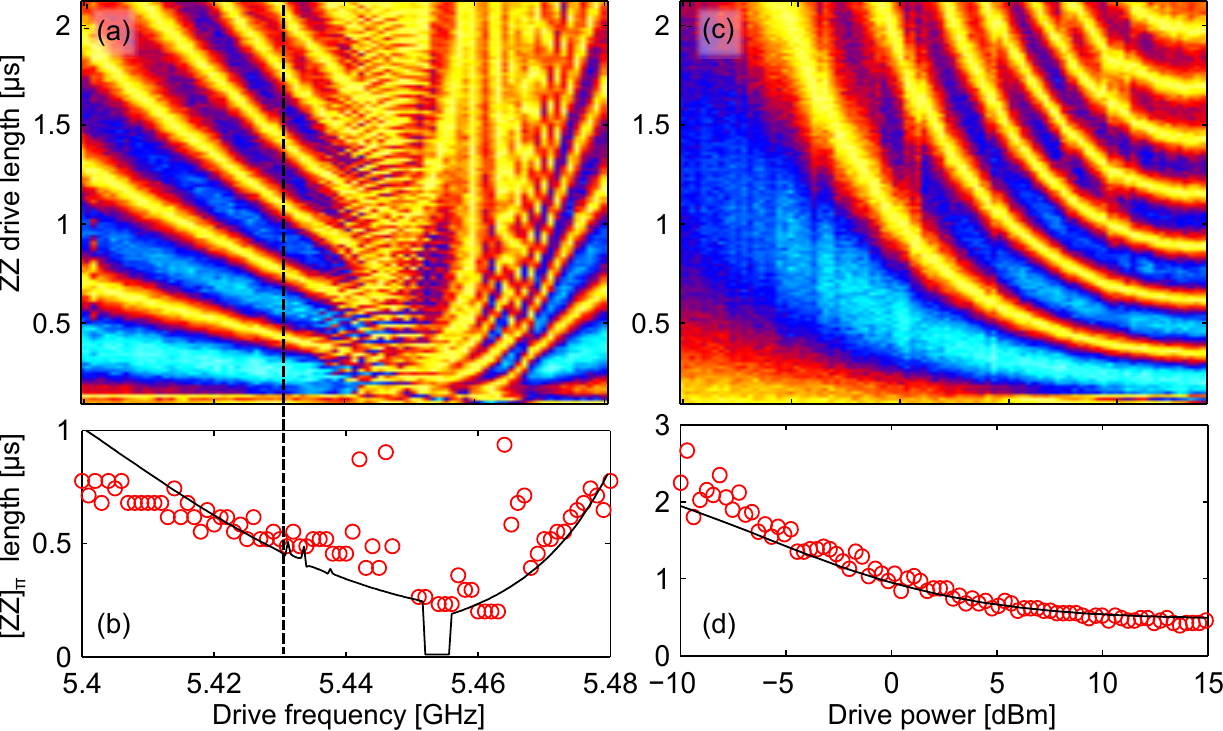}
\caption{\label{fig:3}~(color online) MAP interaction versus frequency and power. (a)~Phase difference between modified MAP scheme Ramsey experiments starting in $\ket{00}$ and $\ket{10}$ swept versus drive frequency and pulse length, with (b)~extracted minimum $[ZZ]_{\pi}$ two-qubit gate length. Dashed line at 5.43 GHz, located outside leakage region (5.443 to 5.466 GHz) containing transitions to $\ket{12}$ and $\ket{03}$. (c)~Phase difference for modified MAP scheme Ramsey experiments starting in $\ket{00}$ and $\ket{10}$ swept versus drive power at 5.43 GHz. (d)~Extracted minimum $[ZZ]_{\pi}$ two-qubit gate length for (c). Solid black lines are from six-level numerical simulations.}
\end{figure}

The MAP drive $\Omega$ can result in additional control errors to both qubits, particularly phase errors $ZI$ or $IZ$ due to the ac-Stark effect, and leakage to higher-levels of Q2 when $\omega_{\text{d}}/2\pi$ is too close to $f_{12}$ of Q2. These can be mitigated with the modified MAP protocol shown in Fig.~\ref{fig:2}(c), which turns the interaction into a two-qubit Clifford generator~\cite{Corcoles2012arx}. In this protocol, the MAP drive is split in half, sandwiched around refocusing $X_{\pi}$ gates on both qubits, and applied only in the CR-like fashion to Q1. The refocusing pulses remove single-qubit phase errors and the CR-like driving of Q1 gives additional protection from leakage to higher levels of Q2. We observe the effect of different Q1 input states on the output of Q2 by ending the Ramsey sequence with a $Y_{\pi/2}$, which results in the oscillations shown in Fig~\ref{fig:2}(d). Here, a $[ZZ]_{\pi}$ is indicated by the dashed line at a total gate time (two MAP drives of 235 ns and single-qubit gate of 40 ns) of 510 ns. The contrast reduction and upward drift in signal is likely due to relaxation, decoherence and higher-order leakage effects. 

We perform the refocused MAP gate scheme at varying $\omega_{\text{d}}/2\pi$ and powers $\Omega^2$ to characterize the interaction. Fig.~\ref{fig:3}(a) shows the phase difference between the Ramsey fringes when starting in $\ket{10}$ and $\ket{00}$, scanning over the drive frequency. We plot the extracted optimal gate time for a $[ZZ]_{\pi}$ versus  $\omega_{\text{d}}/2\pi$ in Fig.~\ref{fig:3}(b), which diverges in the region between 5.443 to 5.466 GHz, due to direct leakage channels to $\ket{12}$ and $\ket{03}$. Nonetheless, the MAP gate is clearly observable at frequencies detuned from this leakage regime. By parking outside this region with a drive frequency of 5.43 GHz, we scan the MAP interaction versus $\Omega^2$ [(Figs.~\ref{fig:3}(c-d)], saturating to a minimum total gate length of 510 ns. 

A six-level numerical simulation is performed, with energy-level frequencies extracted from Rabi amplitude spectroscopy experiments for both qubits. The simulations are shown as solid black lines in Fig.~\ref{fig:3}. We find that these numerical simulations reproduce key features of the experiment. However, the gate time is very dependent on the exact values for the frequencies and only small variations in these values lead to very different predicted gate times. While perturbation theory predicts the behavior at low drive powers, an exact model that explains the gate duration at high powers will be the topic of future investigations. It should also be appreciated that from a control implementation standpoint, the MAP gate is considerably simpler than CR~\cite{Chow2011} or BR~\cite{Poletto2012}, as there is not a stringent requirement that the phase of the drive be particularly locked to other controls in the system.

\begin{figure}[tbp!]
\centering
\includegraphics[width=0.45\textwidth]{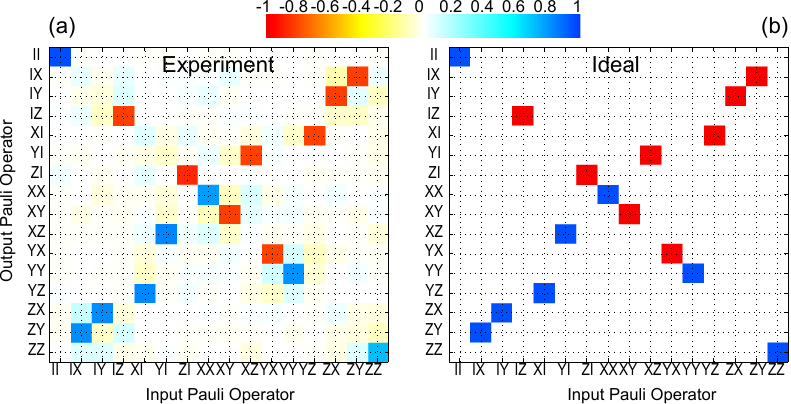}
\caption{\label{fig:4} (color online)~Quantum process tomography for $[ZZ]_{\pi}$ with gate time $\Delta t$ = 510 ns. (a) Experimentally extracted Pauli transfer matrix with gate fidelity $F_{\text{mle}}=0.8717$. (b) Ideal Pauli transfer matrix representation of ideal $[ZZ]_{\pi}$ (with form $(X\otimes X)\text{exp}(-i\frac{\pi}{4}(Z\otimes Z))$).}
\end{figure}

Finally, we perform QPT (Fig.~\ref{fig:4}) by preparing an overcomplete set of 36 states generated by $\{I, X_{\pi},X_{\pm\pi/2},Y_{\pm\pi/2}\}$, applying the 510 ns $[ZZ]_{\pi}$ gate, and performing full state tomography using the bus resonator as a joint readout~\cite{filipp_joint_2009}. We use a semidefinite post-processing algorithm and the Pauli transfer matrix representation~\cite{Chow2012} to represent the process matrix, from which we obtain 
a gate fidelity of $F_{\text{mle}}=0.8717$ [$F_{\text{raw}}=0.876$, $\eta=-0.0285$ (sum of negative eigenvalues of Choi matrix)].
This result is in agreement with the relaxation and decoherence times $(T_1,\,T_2)$ of $(6,\,4)$ $\mu$s for both qubits and the total gate time of 510 ns, likely with some contribution from state preparation and measurement (SPAM) errors~\cite{Chow2012}. SPAM errors can to lead to non-physical effects that make determination of error bars difficult~\cite{Merkel2012arx}.

This experiment represents a proof-of-principle of the MAP gate. Further improvements in gate speed will be possible with more accurate frequency placement of the qubits as well as pulse shaping including amplitude and frequency modulations. Preliminary simulations suggest that gate speeds on the order of $1/J$ should be possible, and clearly further theoretical work is necessary to optimize gate performance.

In conclusion, we have demonstrated a microwave-activated c-Phase gate, utilizing a fixed resonance condition in the higher-energy manifolds of two transmon qubits. With a refocused implementation of the MAP gate, we achieve an optimal $[ZZ]_{\pi}$ in 510 ns and extract gate fidelity of $\approx 87\%$ from QPT. The refocused MAP gate is easily extendable to randomized benchmarking methods~\cite{Magesan2011,Corcoles2012arx} and will be explored in future work. The MAP scheme can be further generalized to any pair of multi-level quantum systems, defining static resonance conditions in non-computational energy states which can be driven to change relative phases on computational states. For superconducting qubits, the MAP scheme is a gate for consideration in larger fixed-frequency quantum processors, but places more stringent boundaries on fabrication.

\begin{acknowledgments}
We acknowledge contributions from A.~D.~C\'orcoles, D.~DiVincenzo, G.~A.~Keefe, J.~Rohrs, M.~B.~Rothwell, and J.~R.~Rozen. We acknowledge support from IARPA under contract W911NF-10-1-0324. All statements of fact, opinion or conclusions contained herein are those of the authors and should not be construed as representing the official views or policies of the U.S. Government.
\end{acknowledgments}

\end{document}